\documentstyle[prd,tighten,aps]{revtex}

\begin{document}

\title {An  effective  Hamiltonian for   2D  black hole Physics}
\author { D. G. Delfrate, F. P. Devecchi \thanks{devecchi@fisica.ufpr.br} and
 D. F. Marchioro }
\address{Departamento de F\'{\i}sica, Universidade Federal
 do Paran\'a, c.p. 19091, cep 81531.990, Curitiba-PR, Brazil.}

\maketitle
\begin{abstract}
In another application of the methods of Henneaux, Teitelboim, and Vergara
 developed for
diffeomorphisms invariant models,  the CGHS
theory  of 2D black holes is focused in order to obtain the true degrees of
 freedom,
the simplectic structure and the {\it effective} Hamiltonian
that rules the dynamics in reduced phase-space.
\end{abstract}

\section {Introduction}
The so-called ``zero-Hamiltonian problem" (ZH problem) is present
in diffeomorphisms invariant models \cite{Dir}.  Being more specific, we
 can say that in
theories of gravitation the canonical hamiltonian density is a linear
combination of constraints;  therefore, after complete gauge fixing,
it reduces to a {\it strongly zero} quantity. 
The ZH problem was analysed by Henneaux, Teitelboim and Vergara
\cite {Hen}:  the idea was to construct an extension on the original
action that is invariant under gauge transformations not vanishing at the
end-points; the boundary conditions were then modified through the gauge
generators. 
The extension mentioned above is related to the physical ({\it effective})
Hamiltonian of the theory that is going to rule the dynamics of the physical
 degrees of freedom.
An alternative approach was proposed by Fulop, Gitman and Tyutin
\cite {Gitman}; the main point here is  that one works in the reduced
 phase-space.
 Once
determined the simplectic structure, after  {\it complete} gauge fixing,
 an special time-dependent
canonical transformation is performed, obtaining  the generator of dynamics
for the physical variables.
 A direct application of these         techniques       was done 
in the 2D {\it induced} gravity model of Polyakov \cite{Pol}; 
here \cite{Deve}
it was possible to obtain the physical Hamiltonian and dynamics of
the true degrees of freedom in a systematic way; without the complications
found when  other  methods are used \cite{Deve2}.
The ZH  problem  was also  focused in the 
2D black hole theory, using {\it dilatonic} gravity models \cite{Bilal}
 (in particular, the CGHS model). Here a different approach was considered:
 the basic idea was  that the
physical Hamiltonian must be a {\it proper} quantity \cite{Regge}. Although
the results were consistent the
techniques used were  not systematical, using several arbitrary
assumptions related to the particular model under analysis. In this work,
ss an  application of the methods of Henneaux et al. and Gitman et al., we
focus
  the 2D CGHS gravity
model\cite{Stro} and show that it is possible to recognize its true degrees
of freedom and obtain the correspondent  reduced
 phase-space physics in a step-by-step
procedure; the key point is the calculation of the {\it effective}
hamiltonian density using those methods.
The manuscript is structured as follows. In the second section  we make a
brief description, as a review, of the techniques used to
 analyse the
ZH problem  in the {\it induced} gravity case.
In the third section  we  follow the Henneaux et al. method to study  the
CGHS model, working in the conformal gauge as a concrete example.
 In the final section  we display
our conclusions.                 

\section {The ZH problem in induced 2D gravity }
The  presence of the ZH problem in field theories is a consequence of
diffeomorphisms invariance \cite{Dir}\cite{Ber}. As is well known, in
those  systems the extended hamiltonian density  ($H_E$) is a linear combination
 of  constraints
\begin{equation}
H_E = H_0+\lambda ^a G_a\approx 0\,\,\, , \label{3}
\end{equation}
\noindent where  $G_a$ represent the first-class constraints. Therefore,
 the Hamiltonian  is a { strongly zero} quantity after the
(complete) gauge fixing procedure,  leaving  no generator of dynamics
in  reduced phase-space \cite{Dir}.  Henneaux, Teitelboim and Vergara
 \cite{Hen}
proposed to perform an extension on the action that takes into account 
 end-point contributions.
The action for the paths obeying these open boundary conditions
(the gauge  parameters $\epsilon ^a$ do not vanish at the end points) is
\begin{equation}
S = \int _{\tau _1}^{\tau _2}( pq-H_0-\lambda ^a G_a )
 d\tau -[P_i\frac{\partial G}{\partial P_i}-G]^{\tau _2}_{\tau _1}\,\,\, ,
\label{4}
\end{equation}
\noindent with $G\equiv \epsilon ^aG_a$.
The corresponding  {\it generating function} ($M$)
is related to the gauge ({\it Diff }) generator. We have in fact
\begin{equation}
M=P_i\frac{\partial G}{\partial P_i}-G\,\,\, .
\label{6}
\end{equation} 
In the {\it induced} 2D gravity case\cite{Pol} the action is given by
\begin{equation}
S=\int d^2x \sqrt{-g} \left(-\varphi \nabla ^{\mu }\nabla _{\mu} \varphi -
\alpha R\varphi
\right) \,\,\, ,\label{50}
\end{equation}
\noindent where $\varphi (x)$ is an auxiliary field and $R$ is the 2D scalar
curvature. For this model the generating function $M$ is \cite{Deve}
\begin{equation}
M = \int  dy \, m
=  \int dy \, [P^i \frac{ \partial G}{\partial P^i}-G] \label{59} \,\,\, ,
\end{equation}
\noindent with

\begin{center}
\begin{eqnarray}
m =   \frac {\sqrt {-g}}{g_{11}}
\epsilon ^0 \left[ \frac{\varphi ^2 }{2} + \frac{2}{\alpha ^2} (g_{11}
\pi ^{11})^2-\frac{\alpha \partial _1 g_{11}}{2g_{11}}
\partial _1 \varphi  \right.\nonumber
\\ \left. + \alpha \partial _1 ^2 \varphi
+ \frac{2}{\alpha } g_{11} \pi ^{11} \pi _{\varphi }\right]
+ (\epsilon ^1 + \frac {g_{01}}{g_{11}}\epsilon ^0 )\left[
-2 g_{11}\partial _1 \pi ^{11}-2\partial _1 g_{11}\pi ^{11}\right]
\label{20} \,\, \, ,
\end{eqnarray}
\end{center}
\noindent being a function of the gravitational field components 
$g_{\mu \nu }$,
the auxiliary field $\varphi $ and their conjugated momenta.
It is also possible to obtain a  non-zero {\it effective} Hamiltonian
($\bar H$), in reduced phase-space, using the technique proposed in
\cite{Gitman}. After complete gauge fixing
a  canonical transformation is performed, whose generator ($F$)
is  determined by the form
of the gauge fixing constraints \cite{Gitman}:

\begin{equation}
\bar H= \left[ H_E + \frac {\partial F}{\partial \tau } \right]
 \vert  _{fixed} \,\,\, .
\label{8}
\end{equation}

 \noindent We obtain the last equality  after (complete) gauge fixing,
 meaning that $\bar H$ is the
Hamiltonian  for the new variables ($\bar Q$), in reduced phase-space.
 The equations
 of motion will be

\begin{equation}
\dot {\bar Q} =\{ \bar Q, \bar H\} _D\,\,\,\,\, \dot {\bar P}=
\{ \bar P, \bar H \} _D\,\,\, ,
\label{9}
\end{equation}

\noindent where D denotes the Dirac bracket \cite {Dir} operation.
In the 2D induced gravity case we found  \cite{Deve}

\begin{equation}
\{g_{11} (x), \pi ^{11} (y) \}_D= \delta (x-y)\,\,\, . \label{34}
\end{equation}

\noindent  Although this is the canonical bracket
relation the gravitational field and the corresponding momentum are not
 independent
quantities in this case \cite{Deve}. 
 For the   {\it effective} hamiltonian density we obtained
\cite{Deve}

\begin{equation}
H_{eff}=g_{11}+ \alpha \left( 1-\frac{1+g_{11}}{2g_{11}} \right)
\pi ^{11} \partial _1 g_{11}\,\,\, ,\label{39}
\end{equation}

\noindent      this density rules the dynamics of the physical gravitational
field ($g_{11}$) in  reduced
 phase-space.

\section {The  2D black hole }

2D black hole physics can be described using the CGHS model of {\it 
dilatonic} gravity \cite {Stro}.
The corresponding action can be written in the following form

\begin{equation}
S=\int d^2x \sqrt{-g} \left(\eta  R - \lambda \right) \,\,\, ,\label{51}
\end{equation}

\noindent  the cosmological constant
is $\lambda$; $\eta$ is related to  the dilaton field $ \varphi $
through $
\eta = e^ {-2\varphi}
$ ; in
 the following we will call $\eta $  the dilaton field for simplification.
 The 2D scalar curvature $R$ is constructed
out of the metric $g_{\mu \nu}$ as usual, while the physical
 gravitational field is
 represented by the $\bar g_{\mu \nu}=\frac{g_{\mu \nu}}{\eta}$  components.
Being a gravitation theory the local gauge
 transformations are the 2D diffeomorphisms. In fact,
  The dilaton  and gravitational
fields transform as

\label{52}
\begin{eqnarray}
\delta \eta &=& \epsilon ^{\mu } \partial _{\mu } \eta \\
\delta g^{ \mu \nu } &=& \partial _{\sigma }
g^{\mu \nu } \epsilon ^{\sigma } - g^{\mu \sigma }
\partial _{\sigma }\epsilon ^{\nu }
+g^{\nu \sigma } \partial _{\sigma } \epsilon ^{\mu }\,\,\, .
\end{eqnarray}

\noindent Using  the physical  gravitational field components
 we can put the action (\ref{50})
in a more convenient form

\begin{equation}
\label{53}
\bar{I}_2 = \int d^2x\,\sqrt{-\bar{g}}\, e^{-2\varphi}\left( \bar{R} + 4
\bar{g}^{\mu\nu} \partial_\mu\varphi\partial_\nu \varphi - \Lambda\right)
\,\,\, .
\end{equation}

Using a new set of  canonical transformations it is possible
 to write the metric components in terms of the shift
vector  $N$ and the lapse function $n$. We have

\begin{equation}
\label{54}
\phi = \frac{1}{4\alpha}e^{-2\varphi} \,\,\, , \,\,\,
\tilde g_{\mu \nu} = \frac{1}{4\alpha \phi} e^{\frac {\phi}{\alpha} }
\bar g_{\mu \nu}\, ,
\end{equation}

\begin{equation}
\label{55}
\tilde g_{00} =  -N^2 + n^2g \label{mlett:1} \,\,\, , \,\,\,
\tilde g_{01} = ng  \,\,\, , \,\,\,
\tilde g_{11} = g \,\,\, .
\end{equation}

\noindent With these new  fields we furnish the primary (first class)
constraints
 of the theory namely $ \pi _{N}=0$ and $ \pi _{n}=0$. In turn, the
 consistency-in-time condition of these quantities give two secondary
 constraints

\label{56}
\begin{eqnarray}
  \omega _1  &=& \frac{1}{2} \left( {\partial _1\phi} ^{2}-\frac{4}{\alpha
^2}(g
\pi _{g})^2-
\frac{4}{\alpha }(g\pi _{g})\pi _{\phi}
  -  \alpha
\frac {\partial _1 g}{
g}+2\alpha \partial _1 ^2 \phi  + \alpha ^2\beta g\right)
\label{mlett:1}\\
\omega_2  &=& \pi _{\phi }
\partial _1\phi  -2g \partial _1\pi _{g}-\pi _{g} \partial _1 g\,\,
 \label{mlett:2},
\end{eqnarray}

\noindent and the canonical hamiltonian density is, as expected, a
 combination of
these secondary (first class) constraints

\begin{equation}
H_c= \frac{N}{g} \omega _1 + n \omega _2 \,\,\, . \label{60}
\end{equation}

In an analogous procedure to the one used in \cite{Deve} we choose
 as gauge fixing conditions

\begin{equation}
\Gamma _5=\pi _{g}-f(t) \,\,\,\,\,\,\, \Gamma _6 =\partial _1 \phi -1
\label{61} \,\,\, .
\end{equation}                    

\noindent where $f(t)$ is an arbitrary function of time.
 To obtain a more convenient form of the Dirac matrix we use the following
 linear
combinations

\begin{equation}
\label{62}
\Lambda _1 =  \omega _1 +\Gamma _5 \,\,\,\, , \,\,\,
\Lambda _2 = \omega _1-\Gamma _5  \,\,\, ,
\end{equation}

\noindent
whose  Poisson brackets are

\begin{eqnarray}
\label{63}
\{\Lambda _1(x), \Lambda _1(y)\}=-2\alpha \partial _x \delta
(x-y)\,\,\, .
\end{eqnarray}

The Dirac brackets for the physical degrees of freedom can be obtained in a
two-steps procedure. First we fix the $[\pi _{n}$ , $\pi _{N}]$ sector
 using the conformal gauge fixing condition
\cite{Bilal}; this is straightforward. In a second step we take the
sector formed by $\omega _1$, $\omega _2$, $\Lambda _1$
and $\Lambda _2$.
We obtain as fundamental Dirac bracket

\begin{equation}
\{g (x), \pi _g  (y) \}_D= \delta (x-y)\,\,\, , \label{64}
\end{equation}

\noindent analogously to the {\it induced} gravity case; the complete
simplectic structure  of reduced phase-space follows from this relation.
 To find the {\it effective} Hamiltonian in reduced phase-space
 we perform, as
 was explained in section II, a time-dependent  canonical transformation.
 In the  gravity sector  we have

 \begin{equation}
\label{66}
\Pi _g = \pi _g  - f(t) \,\,\,\, , \,\,\,\,
G = g \,\,\, .
\end{equation}

\noindent The new Lagrangian density reads

\begin{equation}
\bar L = L + \partial _{\mu } F^{\mu } \,\,\, , \label{67}
\end{equation}

\noindent where $F^{\mu}$ is the generator of the
 canonical transformation. The
correct equations of motion are obtained when

\begin{equation}
\label{68}
F^0=\Pi  g  \,\,\,\, , \,\,\, F^1= \alpha \left(
1-\frac{1+g}{2g}
\right) g' \Pi _g \,\,\, .
\end{equation}

\noindent  Going back to the ``physical''  variables of action (\ref{53})
 we finally obtain that
the {\it effective} hamiltonian density is

\begin{equation}
H_{eff}=\frac{1}{2\pi}\left( e^{-2\varphi } [2 \partial _1 \varphi +
\lambda ] \right)
 \,\,\, .  \label{69}
\end{equation}

\noindent This is the ADM hamiltonian density,
 in conformal gauge, in agreement with  \cite{Bilal}; the dilaton
 appears as the fundamental field (in fact, $g$ is related to the dilaton
 through
the secondary constraints $\omega _1$ and $\omega _2$).  
 The hamiltonian density  (\ref{69}) replaces the original ( equation
(\ref{60}), that is strongly zero after complete gauge fixing) for  the
reproduction of
 the black hole equations, using the simplectic structure of
the reduced phase-space.

\section{Conclusions}

Using the methods developed by Henneaux, Teitelboim and Vergara;
 Gitman and Tyutin, we have obtained  a  solution to the
  ZH problem
 for the  2D  {\it dilatonic}  gravity model, getting  the
 reduced phase-space Physics whose hamiltonian density is in
 agreement  with the one
found in \cite{Bilal} (for conformal gauge fixing). The expressions are the
result of a totally systematic approach; contrary to what is usually  found
 in the literature when considering diffeomorphisms invariant models.

{\bf Acknowledgements}

The authors would like to thank  FUNPAR and PIBIC/CNPq for financial
 support.

\end{document}